# Faraday Cup Measurements of Ions Backstreaming into a Electron Beam Impinging on a Plasma Plume

G. Guethlein, T. Houck, J. McCarrick, and S. Sampayan, LLNL, Livermore, CA 94550, USA

*Abstract*

The next generation of radiographic machines based on induction accelerators is expected to generate multiple, small diameter x-ray spots of high intensity. Experiments to study the interaction of the electron beam with plasmas generated at the x-ray converter and at beamline septa are being performed at the Lawrence Livermore National Laboratory (LLNL) using the 6-MeV, 2-kA Experimental Test Accelerator (ETA) electron beam. The physics issues of concern can be separated into two categories. The interaction of subsequent beam pulses with the expanding plasma plume generated by earlier pulses striking the x-ray converter or a septum, and the more subtle effect involving the extraction of light ions from a plasma by the head of the beam pulse. These light ions may be due to contaminants on the surface of the beam pipe or converter, or, for subsequent pulses, in the material of the converter. The space charge depression of the beam could accelerate the light ions to velocities of several mm/ns. As the ions moved through the body of the incoming pulse, the beam would be pinched resulting in a moving focus prior to the converter and a time varying x-ray spot. Studies of the beam-generated plasma at the x-ray converter have been previously reported. In this paper we describe Faraday cup measurements performed to detect and quantify the flow of backstreaming ions as the ETA beam pulse impinges on preformed plasma.

## 1 INTRODUCTION

The interaction of an intense electron beam with the x-ray converter in radiographic machines is an active area of research[1]. A small, stable (constant diameter and position) electron beam spot size on the converter is essential to achieving a high-quality radiograph. Beam parameters such as emittance and energy variation have been considered limiting factors for realizing the optimum spot size. However, advancements in induction accelerator technology have improved beam quality to a level where the beam interaction with the converter may be the limitation for the next generation of radiographic machines. Two areas of concern are the emission of light ions [2] that can "backstream" through the beam due to space charge potential, and interaction between the beam and the plasma generated by previous pulses during multiple pulse operation.

Previously reported studies have described the use of Faraday cups to characterize the plasma plume generated by the beam at the x-ray converter[3]. We have now used Faraday cups to detect ions that are extracted from a plasma plume by the electron beam and "backstream" through the beam. The studies reported below were performed on the ETA-II accelerator at LLNL using a 6–MeV, 2-kA, 70-ns electron beam. The plasma plume was generated by either a laser or a flashboard.

## 2 EXPERIMENTAL LAYOUT

### 2.1 Faraday Cups

The Faraday cups were comprised of two, electrically isolated, concentric cylinders as illustrated in Fig. 1. The inner cylinder could be biased up to 1.2 kV with respect to the grounded outer cylinder. The OD was 5 cm with an aperture of 1.9 cm. The low ratio of aperture to cup length was to minimize the escape of secondary electrons generated by the impact of the positive ions with the inner cylinder. As shown in Fig. 2, the cups were situated at the entrance of a solenoid operating with an on-axis peak field of approximately 5 kG.

The inner cup discharged to ground through the 50-Ω input of an oscilloscope, permitting the rate of charge interception (current) to be measured. The sensitivity of the cups to ion density, assuming single ionization, is:

$$n_{min} = \frac{I_{min}}{Aev} \text{ , where} \qquad (1)$$

$n_{min}$ is the minimum density, $I_{min}$ is the minimum detectable current, $A$ is the aperture area, and $v$ is the ion velocity. For a nominal $v$ of 2 mm/ns, $n_{min}$ is $2 \times 10^5$ cm$^{-3}$ ($I_{min}$ was 80 μA).

### 2.2 Target Chamber

The x-ray converter was comprised of a rotating wheel that held several "targets" to permit multiple shots before the x-ray converter had to be replaced. The majority of data was taken for tantalum targets of three thicknesses: 1 mm, 0.25 mm, and 0.127 mm. A series of experiments were also performed wherein a thin foil was placed from 5 mm to 15 mm in front of the target to prevent ions produced at the target from backstreaming into the beam. Two-micron thick nitrocellulose and five-micron thick Mylar foils were used. For some runs, Al was sputtered onto the Mylar film to produce a conducting surface.

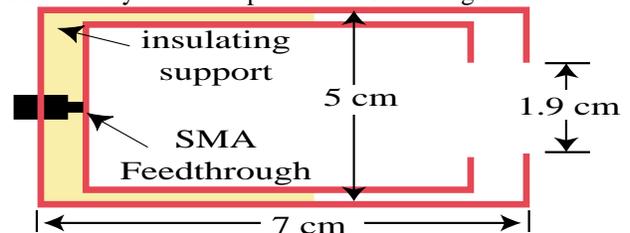

Figure 1: Schematic of the Faraday cup.

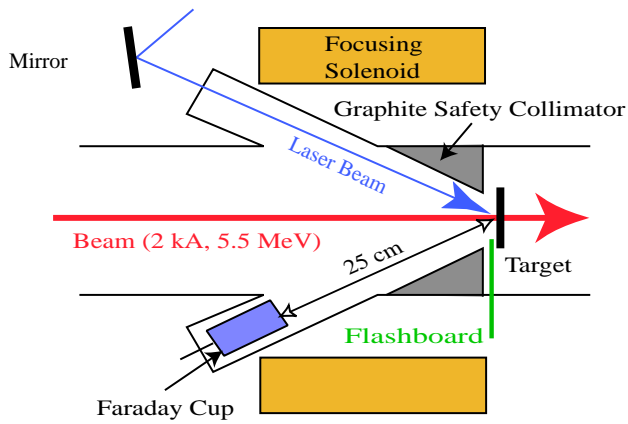

Figure 2: Schematic showing relative positions of the faraday cups with respect to the beam line and target.

Two different target chambers were used. The principal differences between their configurations were the location of the focusing solenoid with respect to the target, the angle of the viewing ports with respect to axis, and number of ports (six or eight) that contained the Faraday cups. The cups were located about 25 cm from the beam/target intersection for both configurations. In the first configuration the cups were at an angle of 30° from the beam axis and located partially under the solenoid. The second configuration reduced the angle to 20° and the cups were just outside the solenoid entrance. Data shown below are for two cups located approximately on opposite sides of the beamline; cup #1 was towards the bottom and cup #2 was at the top.

The actual chambers did not have the perfect cylindrical symmetry shown in Figure 2. Various diagnostics and access ports were located around the chambers. However, the largest deviation from the axial symmetry was the target wheel. The axis of this wheel was located several centimeters below the beam axis to avoid being struck and damaged by the beam. The OD of the wheel extended from about two cm above the axis to about 7 cm below. Some combination of the physical geometry of the system led to azimuthal asymmetries in the Faraday cup data.

### 2.3 Laser and Flashboard

An 0.8J, 10 ns FWHM, Nd:YAG laser was directed at the target and timed to produce a plasma of sufficient density that would simulate target debris such as multiple pulse electron beams might encounter near the target. The pulse energy given below is the energy measured at the laser. The energy at the target was approximately half that value and decreased as the window was covered with debris.

A flashboard was added to the experiment to generate a "cooler" plasma than was possible with the laser. The flashboard was constructed from semi-rigid coax cable by cutting an end so the inner conductor protruded about 3 mm. The inner conductor was flattened and covered with a graphite solution (aerodag) to enhance breakdown to the outer conductor. A 5 kV pulse was applied to the cable producing an arc at the tip that was sustained for a few microseconds. Although relatively simple in construction, the flashboard produced plasma plumes that were more consistent in density and velocity than those generated using the laser. The tip of the flashboard was located about 1.5 cm from the beam/target intersection.

## 3 RESULTS AND DISCUSSION

The first ion signature measured by the Faraday cups occurred when a 250-mJ laser pulse impacted the converter about 30 ns prior to beam time. The output of one of the cups is shown in Figure 3 superimposed on a signal when there was no prebeam laser pulse. The small signal labeled "laser pulse at converter" was due to UV radiation (generated by the laser at the converter) knocking electrons off the surface of the cup. The larger signal labeled "e-Beam Signal" is due to beam electrons scattered off the converter[4]. Both signals are detected at the Faraday cup within 1 ns of the respective events and were used as timing fiducials. The large positive signal following the beam was caused by ions extracted from the laser produced plasma plume. In the 30 ns before beam arrival, this plume would have expanded only a few mm from the converter's surface. Ions were detected for experiments where the laser pulses arrived on target as late as midway through the e-beam pulse.

The species and energy of the ions can be inferred if some assumptions are made about the path the ions took to the cup, the relative time the ions were accelerated into the e-beam, and the magnitude of the e-beam space charge depression. With a number of qualifications, we believe that the ion signal was due primarily to $H^+$ ions with energy on the order of 100 KeV. An order of magnitude estimate of the ion density can be made by assuming that the cup intersected a portion of a uniform $2\pi$-stereradian distribution of ions expanding from the target. For Figure 3, the ion density in the vicinity of the target is about $10^{13}$ cm$^{-3}$.

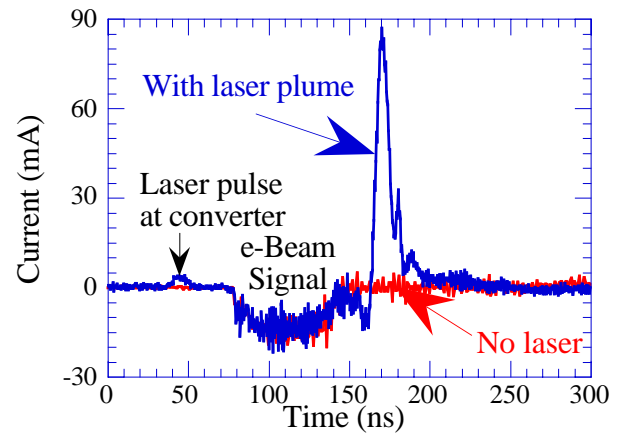

Figure 3: Faraday cup signal for an e-beam striking a laser generated plasma. There was no bias on the cup.

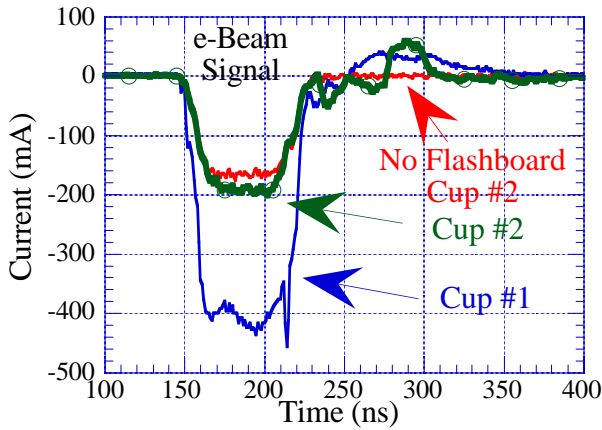

Figure 4: Faraday cup signals for an e-beam striking a flashboard generated plasma.

Figure 4 shows Faraday Cup signals when the beam impacted the plasma plume generated by the flashboard in front of the target superimposed on the baseline Cup #2 signal for no flashboard. The figure also illustrates the difference in signals received at different azimuthal positions. The flashboard was fired about a microsecond before beam arrival to ensure that the plasma plume had reached the beam/target intersection. The later arrival of the main ion signal infers that heavier ions were involved than for the case shown in Figure 3. For this case the ion signal could have been caused by $C^+$ ions with energy on the order of 100 KeV in addition to $H^+$ ions.

A non-conducting, thin film was suspended in front of the target to impede the flow of ions into the beam. We did not believe that sufficient energy would be deposited into the film by the beam to ionize material. The lack of ions generated by the beam on Ta targets supported this scheme. However, as shown in Figure 5, the beam striking the film could generate a large ion signal. This occurred sporadically when the film was 1.5 cm in front of the target, essentially every time at 1.0 cm, and never at 0.5 cm. A conductive coating on the film prevented ion generation. A similar effect was noted on experiments performed on the PIVAIR accelerator at CESTA[5].

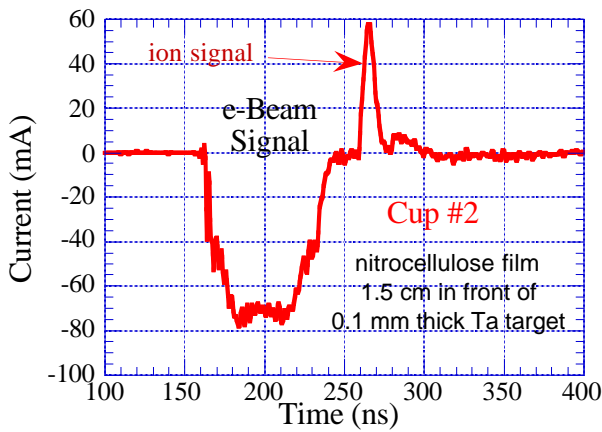

Figure 5: Faraday cup signal for an e-beam striking a non-conducting film 1.5 cm in front of a Ta target.

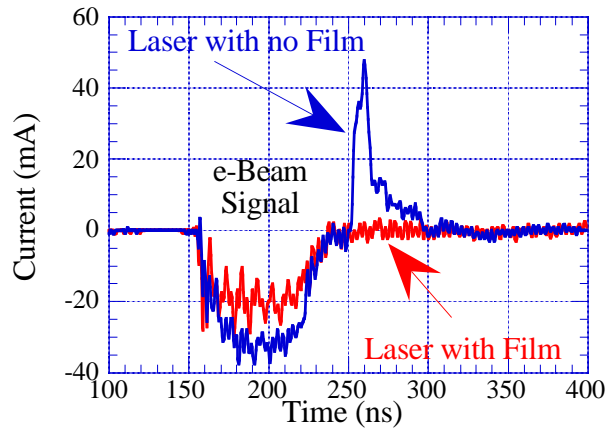

Figure 6: Faraday cup signals showing effect of film on backstreaming ions.

The film was effective at blocking ions in some cases. Figure 6 shows the difference between film and no film when a 53 mJ laser pulse fired 30 ns before the e-beam was used to generate a plasma source for ions. The film was placed 1.5 cm in front of the target for this data.

## 4 SUMMARY

The Faraday cups proved effective at detecting and identifying ions backstreaming into the e-beam.. Indications of these ions were also supported by a time varying beam spot on target as observed by the x-ray spot produced. No ions were detected for the e-beam impinging on a metallic target without a preformed plasma.

## 5 ACKNOWLEDGEMENTS

## REFERENCES

[1] Sampayan, S., et al., "Beam-Target Interaction Experiments for Bremsstrahlung Converter Applications," Proc. 1999 Part. Accel. Conf., p. 1303.
[2] G. Caporaso and Chen, Y-J, "Analytic Model of Ion Emission From the Focus of an Intense Relativistic Electron Beam on a Target," Proc. XIX Int'l LINAC Conf., p. 830 (1998).
[3] T. Houck, et al., "Faraday Cup Measurements of the Plasma Plume Produced at an X-Ray Converter," Proc. XIX Int'l LINAC Conf., p. 311 (1998).
[4] Falabella,S., et al., "Effect of Backscattered Electrons on Electron Beam Focus," this conf., TUB11.
[5] C. Vermare, et al., *IEEE Trans. Plasma Sci.*, Vol. 27, No. 6, pp. 1566–1571 Dec. 1999.